\documentclass[12pt]{article}
\usepackage{fullpage}
\usepackage{amsmath}
\usepackage{amssymb}
\usepackage[dvips]{epsfig}

\def\##1{\underline{#1}}
\def\=#1{\underline{\underline{#1}}}

\def\+#1{\underline{\bf #1}}
\def\*#1{\underline{\underline{\bf #1}}}

\def\r#1{(\ref{#1})}
\def\l#1{\label{#1}}
\def\c#1{\cite{#1}}

\def\le{\left(}
\def\ri{\right)}
\def\les{\left[}
\def\ris{\right]}

\def\ric{\right\}}

\def\.{\mbox{ \tiny{$^\bullet$} }}

\def\epso{\epsilon_{\scriptscriptstyle 0}}

\def\muo{\mu_{\scriptscriptstyle 0}}

\def\tr{(ct,\#r)}
\def\ok{(\omega/c,\#k)}

\def\mrbh{m_{\mbox{\tiny{rbh}}}}
\def\arbh{a_{\mbox{\tiny{rbh}}}}

\begin{document}

\begin{center}

{\bf {\Large A comparison of superradiance and negative--phase--velocity phenomenons in
the ergosphere of a rotating black hole }}

 \vspace{10mm} \large

Sandi Setiawan\footnote{Fax: + 44 131 650 6553; e--mail:
S.Setiawan@ed.ac.uk.}, \\
{\em School of Mathematics,
University of Edinburgh, Edinburgh EH9 3JZ, UK}\\
\bigskip
 Tom G. Mackay\footnote{Corresponding Author.
Fax: + 44 131 650 6553; e--mail: T.Mackay@ed.ac.uk; permanent
address: School of Mathematics, University of Edinburgh, Edinburgh
EH9 3JZ, UK},
 Akhlesh  Lakhtakia\footnote{Fax: + 1 814 863
4319; e--mail: akhlesh@psu.edu; also
 affiliated with Department of Physics, Imperial College, London SW7 2 BZ,
UK}\\
 {\em Department of Engineering Science and
Mechanics\\ Pennsylvania State University, University Park, PA
16802--6812, USA}\\

\end{center}

\vspace{4mm}

\normalsize

\begin{abstract}
\noindent The propagation of electromagnetic plane waves with
negative phase velocity (NPV) in the ergosphere of a rotating
black hole has recently been reported. A comparison of NPV
propagation and superradiance is presented. We show
that, although both phenomenons involve negative energy densities,
there are two significant differences between them.

\end{abstract}

\noindent {\bf Keywords:} Negative phase velocity, Superradiance,
Poynting vector, Kerr spacetime

\vspace{10mm}

\section{Introduction}

The notion of negative energy is often regarded with suspicion.
 However, negative energy densities themselves are not
 uncommon, even within the realm of classical physics \c{BV00}. A
 straightforward
 example is  provided by Newton's law of gravitation.
  From the perspective of an observer at infinity, negative energy
is associated with bound states in the Newtonian gravitational
field. Similarly, negative energy is associated with  bound states
in electrostatics \c{HRW}.

The concept of negative energy lies at the very heart of  quantum
mechanics, courtesy of the uncertainty principle. Vacuum
fluctuations give rise to manifestations of negative energy
density. For example, the so--called squeezed states of light in
quantum optics are characterized by  periodic spatial
distributions
 of positive and negative energy density \c{Slusher,Ford}.
Another example is furnished by the Casimir effect, which is a
negative--energy  phenomenon existing between two closely spaced
parallel plates \c{Spruch,Lamoreaux,Barton}.

Negative energy densities are particularly associated with
astrophysical settings. Notably, the emission of Hawking radiation
by a black hole is accompanied by a flow of negative energy into
the black hole \c{Hawking}.  Also, the construction of
wormholes~---~i.e., hypothesized tunnels linking regions of curved
spacetime~---~relies on negative energy density \c{BV00,Safonova}.

It is widely known that the region of spacetime immediately
surrounding the event horizon of a rotating black hole, namely the
ergosphere, supports the negative--energy phenomenon of
superradiance \c{Chandra,Schutz}. By this process, the
extraction of energy from the black hole~---~via the creation of
negative--energy photons within the ergosphere~---~may be
envisaged. Recently, we reported on an unusual form of
electromagnetic planewave propagation, called negative phase
velocity (NPV) propagation,  in the ergosphere of the Kerr black
hole \c{LMS05_PLA,MLS05_CQG}. NPV propagation also appears to
involve  negative energy density \c{MLS05_NJP}. The question naturally arises:
are superradiance and NPV propagation related? We
address this question in the following sections.

\section{Superradiance}

Let us briefly review the  salient features of black--hole
superradiance; comprehensive descriptions can be found elsewhere
\c{Chandra,Schutz}. We consider the rotating black hole described
by the Kerr metric.  In Boyer--Lindquist coordinates, the line
element in Kerr spacetime is expressed as
\begin{eqnarray}
ds^2 &=& \frac{\Delta}{\rho^2} \le dt - \arbh \,
 \sin^2 \theta \, d\phi \ri^2 - \frac{\sin^2 \theta}{\rho^2}
  \les \le R^2+\arbh^2 \ri d\phi - \arbh \, dt\ris^2 \nonumber \\
&& - \frac{\rho^2}{\Delta} \, dR^2 - \rho^2 \, d\theta^2 \,,
\end{eqnarray}
with
\begin{eqnarray}
\Delta &=& R^2 - 2\mrbh R + \arbh^2, \\
\rho^2 &=& R^2 + \arbh^2 \, \cos^2 \theta \, .
\end{eqnarray}
Herein the metric signature $(+,-,-,-)$ is adopted for a black
hole of geometric mass $\mrbh$. The term $\arbh$  is a measure of
the black hole's angular velocity.
 The coordinate $\phi$ is the azimuthal angle around the
axis of rotation~---~which is taken as the $z$ axis of a Cartesian
coordinate system~---~and $t$ is the time coordinate. The
coordinate $R$ is related to the Cartesian coordinates ($x$, $y$,
$z$) via
\begin{equation}
R^2 = x^2 + y^2 + z^2 - \arbh^2 \le 1 - \cos^2 \theta \ri
\end{equation}
with $\cos \theta = z/R$.

The outer solution to $\Delta = 0$ corresponds to the outer event
horizon of the Kerr black hole; it is denoted by $R=R_+$ where
\begin{equation}
R_+ = \mrbh + \sqrt{\mrbh^2 - \arbh^2} \,,
\end{equation}
with $ \mrbh^2 >  \arbh^2$.

For simplicity, let us restrict our attention to the equatorial
plane (i.e., $\theta = \pi/2$). The trajectories of photons
initially travelling in the $\pm \phi$ direction are provided by
\begin{equation} \l{photon_eqn}
\frac{d\phi}{dt} = \frac{g_{t\phi}}{g_{\phi\phi}} \pm \sqrt{\le
\frac{g_{t\phi}}{g_{\phi\phi}} \ri^2 -
\frac{g_{tt}}{g_{\phi\phi}}} \, ,
\end{equation}
where $g_{\alpha \beta}$ are the components of the Kerr
metric. Parenthetically, we remark
that a central feature of the Kerr black hole is the the
so--called dragging of inertial frames which results from
   the  off--diagonal  metric component
$g_{t\phi}$ \c{barpet}.

From \r{photon_eqn} we see that  the two solutions
\begin{equation} \l{solns}
\frac{d\phi}{dt} = \left\{\begin{array}{l}
 \displaystyle{ \frac{2 g_{t\phi}}{g_{\phi\phi}}}
\\ \\ 0
\end{array}
\right.
 \, ,
\end{equation}
emerge for $g_{tt} = 0$.
 The nonzero solution
corresponds to a photon initially travelling in the same
directional sense as  the black--hole rotation, whereas the zero
solution corresponds to a photon initially directed in the
opposite sense to the black--hole rotation. Thus, we see that  at
$g_{tt} = 0$ the frame dragging is sufficiently strong that photon
trajectories in the opposite  sense to the black--hole rotation
are not permitted. The surface where $g_{tt} = 0$ is called the
stationary limit surface and it lies at
 $R= R_{S_+}$ where
\begin{equation}
R_{S_+}  = \mrbh + \sqrt{ \mrbh^2 - \le \, \frac{\arbh z}{R_{S_+}}
\,\ri^2}\,.
\end{equation}
The ergosphere is
 defined to be the region between
the outer event horizon and the stationary limit surface; i.e.,
the region specified by $R_+ < R < R_{s_+}$ .

Since the Kerr spacetime is stationary and axisymmetric,
trajectories may be characterized by the  following two
quantities:
\begin{itemize}
\item[(i)] $E$, the energy measured at infinity \c{andersson}, and
\item[(ii)]
$L$, the component
of angular momentum parallel to the symmetry axis.
\end{itemize}
The radial
motion of photons travelling in the equatorial plane is described
by
\begin{equation}
\le \frac{dR}{d\sigma} \ri^2 = \frac{\le R^2+ \arbh^2 \ri^2 -
\arbh^2 \Delta}{R^4} \le E-V_+ \ri \le E-V_- \ri,
\end{equation}
with
\begin{equation}
V_{\pm} = \les \frac{2 \arbh \mrbh R \pm R^2 \Delta^{1/2}}{\le
R^2+ \arbh^2 \ri^2 - \arbh^2 \Delta} \ris L \, ,
\end{equation}
and $\sigma$ being an arbitrary parameter on the photon
trajectory. It follows thereby that, within the ergosphere, photon
trajectories for $\arbh L
> 0$ are characterized by $E > 0$, whereas $E<0$ solutions exist
for $\arbh L < 0$. That is, the ergosphere supports
negative--energy photons provided that their angular momentum is
initially directed opposite to the angular momentum of the black
hole.

The term superradiance is used to describe the spontaneous
emission of positive--energy photons which can take place in the
ergosphere of a Kerr black hole \c{Bekenstein}. If energy is to be
conserved during the emission process \c{Spruch}, the positive energy of the superradiant photons must be
balanced by the creation of negative--energy photons.

\section{Negative--phase--velocity propagation}

The phase velocity of a plane wave is called negative if the
wavevector $\#k$ is directed  opposite  to the time--averaged
Poynting vector $\langle \, \#P \, \, \rangle_t$ \c{LMW}. Thus,
negative--phase--velocity (NPV) propagation is signalled by
\begin{equation}
\#k \. \langle \, \#P \, \, \rangle_t < 0\,. \l{NPV_cond}
\end{equation}
Many interesting consequences follow from \r{NPV_cond}, most
notably the phenomenon of negative refraction \c{Pendry04}. The
technological possibilities offered by negative refraction,
especially relating to the  production of  highly efficient
lenses, has prompted considerable recent interest in artificial
\emph{metamaterials} which support NPV propagation \c{Rama}.
Furthermore, it has been shown that NPV propagation is possible in
vacuum for certain curved spacetime metrics \c{MLS05_NJP}.

A brief review of NPV propagation for  Kerr spacetime
is presented in this section; full details of the analysis are available elsewhere
\c{LMS05_PLA,MLS05_CQG}. Following the standard
approach, first proposed by Tamm \c{Tamm}, electromagnetic
propagation in vacuum for curved spacetime is described in terms
of propagation in the fictitious bianisotropic medium
characterised by the constitutive relations \c{SS}\footnote{Greek indexes
take the values 0, 1, 2, and 3 corresponding to $t$, $x$, $y$, and $z$,
respectively; Roman
indexes take the values 1, 2, and 3.}
\begin{equation}
\label{CR2} \left.
\begin{array}{l}
D_\ell = \gamma_{\ell m} E_m + \varepsilon_{\ell mn}\,\Gamma_m\,H_n\\[6pt]
B_\ell =    \gamma_{\ell m} H_m - \varepsilon_{\ell mn}\,
\Gamma_m\,  E_n
\end{array}\right\}\,.
\end{equation}
Herein, $\varepsilon_{\ell mn}$ is the three--dimensional
Levi--Civita symbol,
\begin{equation}
\label{akh1} \left.\begin{array}{l} \gamma_{\ell m}
= \displaystyle{- \le -{g} \ri^{1/2} \, \frac{{g}^{\ell m}}{{g}_{00}}}\\[6pt]
\Gamma_m= \displaystyle{\frac{g_{0m}}{g_{00}}}
\end{array}\ric
\,,
\end{equation}
and  $g = \mbox{det} \les g_{\alpha \beta} \ris $. More
conveniently, we recast  \r{CR2} in the  conventional 3 $\times$ 3
dyadic/ 3 vector form
\begin{equation}
\left.
\begin{array}{l}
 \label{eq2}
\#D\tr = \epso\,\=\gamma\tr\cdot \#E\tr -
\displaystyle{\frac{1}{c}}\, \#\Gamma\tr\times \#H\tr\,\\
\vspace{-8pt} \\ \#B\tr = \muo\,\=\gamma\tr\cdot \#H\tr +
\displaystyle{\frac{1}{c}}\,\#\Gamma\tr\times \#E\tr\,
\end{array}
\right\},
\end{equation}
where $\=\gamma\tr$ is the dyadic--equivalent of $\gamma_{\ell
m}$,
 $ \#\Gamma\tr$ is the vector--equivalent of $\Gamma_m$;
the scalar constants $\epso$ and $\muo$ denote the permittivity
and permeability of vacuum in the absence of a gravitational
field; $c=1/\sqrt{\epso\muo}$; and SI
 units are adopted.

 Planewave propagation has been investigated within
an arbitrary spacetime neighbourhood $\cal R$ whose spatial
location
 is  given by the
 Cartesian coordinates $\le \tilde{x},
\tilde{y}, \tilde{z} \ri$ \c{MLS05_CQG}. The neighborhood is taken to be
sufficiently small that the nonuniform metric $g_{\alpha \beta}$
may  be approximated by the uniform metric ${\tilde g}_{\alpha
\beta}$ throughout  $\cal R$. Thus, we introduce the uniform
3$\times$3 dyadic $
 \={\tilde\gamma} \equiv \left. \=\gamma \,\right|_{\cal R} $ and
  the uniform 3 vector $ \#{\tilde\Gamma} \equiv \left. \#\Gamma
\, \right|_{\cal R} $. Applying standard techniques of planewave
analysis within the uniform neighbourhood $\cal R$, we previously
derived the expression \c{MLS05_CQG}
\begin{eqnarray}
 \langle \#{\sf P}\rangle_t &=& \frac{1}{2 \omega \muo \vert
\={\tilde\gamma} \vert} \le \vert A_a \vert^2 \#{\sf e}_a \.
\={\tilde\gamma} \. \#{\sf e}_a +  \vert A_b \vert^2 \#{\sf e}_b
\. \={\tilde\gamma} \. \#{\sf e}_b \ri \,  \={\tilde\gamma} \. \#p
\,. \label{ppp1}
\end{eqnarray}
for the time--averaged Poynting vector. The vector $\#p$ in
\r{ppp1} is related to the wavevector $\#k$ through
\begin{equation}
\#p = \#k -  \frac{\omega}{c} \,\#{\tilde\Gamma}\,,
\end{equation}
with $\omega$ being the angular frequency of the plane wave. Furthermore,
the corresponding dispersion relation  yields  two wavenumbers $k
= k^\pm$ for the arbitrarily oriented  $\#k = k \hat{\#k}$ with
$\hat{\#k} = \le \sin \theta \cos \phi, \sin \theta \sin \phi,
\cos \theta \ri$, namely,
\begin{equation}
k^\pm = \frac{\omega}{c} \le \frac{\hat{\#k} \. \={\tilde\gamma}
\. \#{\tilde \Gamma} \pm \sqrt{ \le \, \hat{\#k} \.
\={\tilde\gamma} \. \#{\tilde \Gamma} \, \ri^2 - \hat{\#k} \.
\={\tilde\gamma} \. \hat{\#k} \le \#{\tilde\Gamma} \.
\={\tilde\gamma} \. \#{\tilde\Gamma} - \vert \,\={\tilde\gamma} \,
\vert\, \ri}}{\hat{\#k} \. \={\tilde\gamma} \. \hat{\#k}}\ri\,.
\end{equation}
The complex--valued constants  $ A_{a,b} \ok$, which are fixed by
initial and boundary conditions, provide the amplitudes
accompanying the
 unit eigenvectors
\begin{equation}
\left.
\begin{array}{l}
 \#{\sf e}_a
= \displaystyle{\frac{\={\tilde\gamma}^{-1} \. \#w}
{\vert\={\tilde\gamma}^{-1} \. \#w\vert}} \\ \\\  \#{\sf e}_b =
\displaystyle{\frac{\={\tilde\gamma}^{-1}\. \le \#p \times \#{\sf
e}_a \ri}{\vert\={\tilde\gamma}^{-1}\. \le \#p \times \#{\sf e}_a
\ri\vert}}
\end{array}
\right\}
 \,. \l{e_12}
\end{equation}
The  unit vector $\#w$ is orthogonal to $ \#p$, i.e.,
$\#w\.\#p=0$,  but is otherwise arbitrary.

Since
\begin{equation}
\vert \, \=\gamma \, \vert = \delta^2 \les R^4 + \le \arbh z \ri^2
\ris^2\, > 0\,
\end{equation}
for the Kerr metric, the sufficient conditions
\begin{equation} \l{NPV_cond2}
\left.
\begin{array}{l}
\le \, \#{\sf e}_a \. \={\tilde\gamma} \. \#{\sf e}_a \, \ri \le
\, \#k \. \={\tilde\gamma} \. \#p \, \ri < 0 \\ \vspace{-6pt}
\\
\le \, \#{\sf e}_b \. \={\tilde\gamma} \. \#{\sf e}_b \, \ri \le
\, \#k \. \={\tilde\gamma} \. \#p \, \ri
 < 0
\end{array}
\right\}
\end{equation}
 for NPV propagation follows directly from \r{ppp1}.
In an earlier study, we demonstrated that the NPV sufficient
conditions \r{NPV_cond2} are satisfied for certain wavevector
orientations at various locations throughout the  ergosphere of the Kerr
black hole \c{MLS05_CQG}.

\section{Superradiance versus NPV propagation}

The key feature which is common to both black--hole superradiance
and NPV propagation is negative energy: The electromagnetic energy
density   associated with a NPV plane wave, as observed by an
observer at infinity \c{andersson}, is negative--valued \c{MLS05_NJP,Ruppin}.
Negative--energy photon trajectories are required for black--hole
superradiance. Furthermore, nonrotating black holes (yielding the
Schwarzschild metric by setting $a_{rbh}=0$) support neither superradiance
nor NPV propagation \c{MLS05_CQG}.

However, there are also important differences
between these two phenomenons which we now elaborate on.

\subsection{Angular momentum considerations}

As described earlier,  negative--energy photon trajectories arise
in the Kerr ergosphere only when a photon's angular momentum is
initially directed in the opposite sense to the  black--hole
rotation. Let us investigate how  this compares with the situation
for NPV planewave propagation.

 The  angular momentum density for
a plane wave, at the point $\#r$, is provided by $\#r \times
\langle \, \#P \, \rangle_t$. Furthermore, the component of
angular momentum density parallel to the $z$ axis is given as
$\hat{\#z}\.\le\#r \times \langle \, \#P \, \rangle_t \ri$ where
$\hat{\#z}$ is a unit vector pointing along the positive $z$ axis.
For planewave propagation in the Kerr ergosphere, the sign of
$\hat{\#z}\.\le\#r \times \langle \, \#P \, \rangle_t \ri$ may  be
inferred using
 the expression \r{ppp1} for the time--averaged Poynting vector.
Therefore, it can be deduced whether or not NPV plane waves have
angular momentum parallel or anti--parallel to the black hole
angular momentum.

We present some illustrative numerical results in figures~1 and 2.
For the black hole with the angular velocity term $\arbh =
\sqrt{3/4} \,\mrbh $, we examine three points in the ergosphere,
at locations on the $x$ axis given by $R = 1.55 \,\mrbh$, $R =
1.75 \,\mrbh$ and $R = 1.95 \, \mrbh$. Notice that for $\arbh =
\sqrt{3/4} \, \mrbh $,  the outer event horizon lies at $R_+ = 1.5
\,\mrbh$ whereas the stationary limit surface lies at $R_{S_+} = 2
\, \mrbh$ on the $x$ axis.
 In figure~1 the
orientations of the NPV wavevectors are mapped. It is observed
that the NPV wavevectors generally lie in the equatorial plane.
They  are oriented away from the centre of the black hole for the
$k^+$ wavenumbers and oriented towards the centre of the black
hole for the $k^-$ wavenumbers. Also, we see that the range of
wavevector orientations which support NPV propagation increases as
$R$ decreases.
 The corresponding orientations of the NPV plane wave's
angular momentum relative to $\hat{\#z}$ are mapped in figure~2.
We see that for all of the $k^+$ wavenumbers and some of the $k^-$
wavenumbers, the NPV plane wave's angular momentum is oriented in
the opposite direction to the black hole's  angular momentum. Most
significantly, some of the $k^-$ wavenumbers correspond to NPV
planewave angular momentums which are oriented in the same
direction as the black hole's angular momentum.

Therefore, we conclude that while many NPV modes may be consistent
with the negative--energy photon trajectories of superradiance,
there exist NPV modes which are {\em definitely
 incompatible\/} with the superradiant
scenario.

\subsection{Frequency/wavelength considerations}

Black--hole superradiance is a frequency--bounded phenomenon:
Superradiance occurs only when \c{Chandra,andersson}
\begin{equation}
\omega < m \, \omega_+\,, \l{npv_star}
\end{equation}
where $\omega_+ = \arbh / (2 \mrbh R_+)$ is the angular frequency
associated with the outer event horizon
and $m \geq 1$ is an integer. For spontaneously emitted
photons, the probability of large values of $m$ is small.

In contrast, NPV propagation occurs at short wavelengths with no
upper bound on frequency. To be specific, suppose that the
neighbourhood ${\cal R}$ has representative spatial linear
dimensions given by $\ell$. The approximation of the nonuniform
metric $g_{\alpha \beta}$  by the uniform metric
$\tilde{g}_{\alpha \beta}$ throughout ${\cal R}$ relies upon
$\ell$ being small relative to the radius of curvature of the Kerr
spacetime. The inverse radius of spacetime curvature squared is
conveniently provided by the nonzero components of the Riemann
tensor. For the Kerr metric these components are of the order of
$R^{-2}$  \c{Chandra}. Thus, we have $\ell \lesssim R$. In
addition, the neighbourhood ${\cal R}$  should be large compared
with
 electromagnetic wavelengths, as given by $2 \pi/ |k|$. Hence,
the NPV conditions \r{NPV_cond2}  hold in the regime
\begin{equation} \l{applicability}
\frac{2 \pi}{ | k | } \lesssim  R.
\end{equation}

\section{Concluding remarks}

In the introduction, we posed the question: are black--hole
superradiance and NPV propagation related? We answer
this question as follows. NPV propagation and superradiance
 are related insofar as both involve negative energy
densities within the ergosphere of the  Kerr black hole. However,
there are two significant distinctions to be made.
\begin{itemize}
\item[(a)] The
negative energies in black--hole superradiance are associated with
photons which have angular momentum which is initially oriented in
the opposite sense to the black--hole rotation. In contrast, the
angular momentum of NPV planewaves can be oriented both in the
opposite sense and in the same sense as the black--hole rotation.
\item[(b)] There is practically an upper bound on the frequency of waves which can
undergo  superradiance. In contrast, NPV propagation
occurs at wavelengths which are short relative to the radius of
spacetime curvature, but
 there is
no upper bound on frequency.
\end{itemize}
We conclude by emphasizing that at least two clear differences exist between
the phenomenons of superradiance and NPV propagation in the
ergosphere of a rotating black hole.

\vspace{10mm}

\noindent{\bf Acknowledgements:}

\noindent SS and TGM acknowledge EPSRC for support under grant
GR/S60631/01. TGM thanks the Department of Engineering Science and
Mechanics at Pennsylvania State University for their hospitality.

\vspace{10mm}

\noindent{\bf Note added to proof:}

\noindent The superradiance condition \r{npv_star} was first
established for scalar waves by Starobinksy \c{Star1}, and for
electromagnetic and gravitational waves by Starobinsky and
Churilov \c{Star2}. From the derivations in these papers, it is
clear that superradiance requires the existence of negative energy
photons near the black hole event horizon. The authors thank Prof.
Starobinsky for drawing their attention to this matter.

\newpage

 \setcounter{figure}{0}
\begin{figure}[!ht]
\centering \psfull \epsfig{file=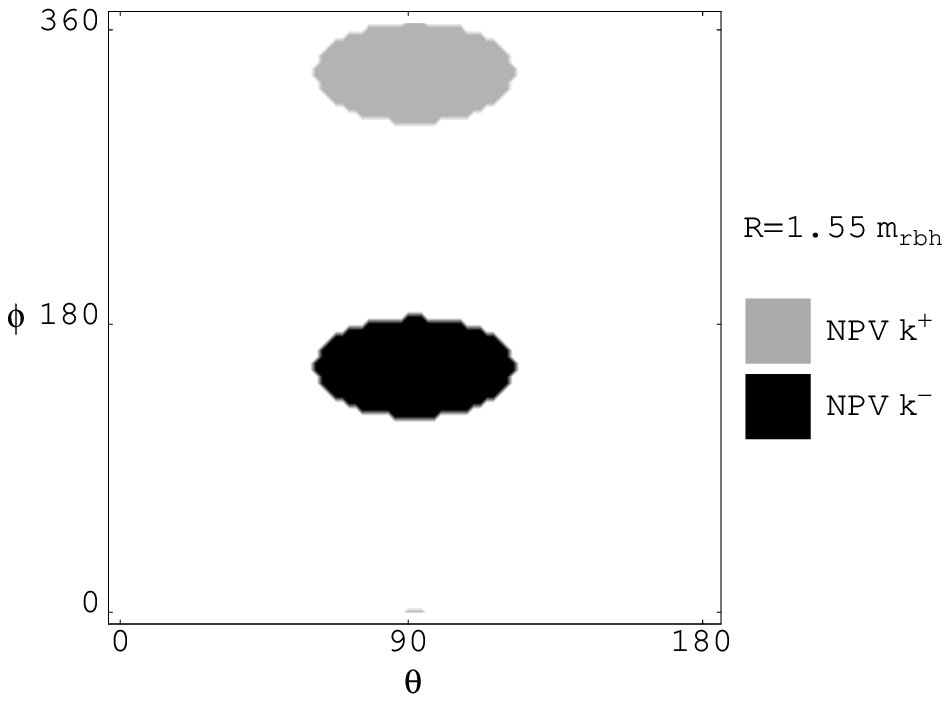,width=3.0in}\\
\epsfig{file=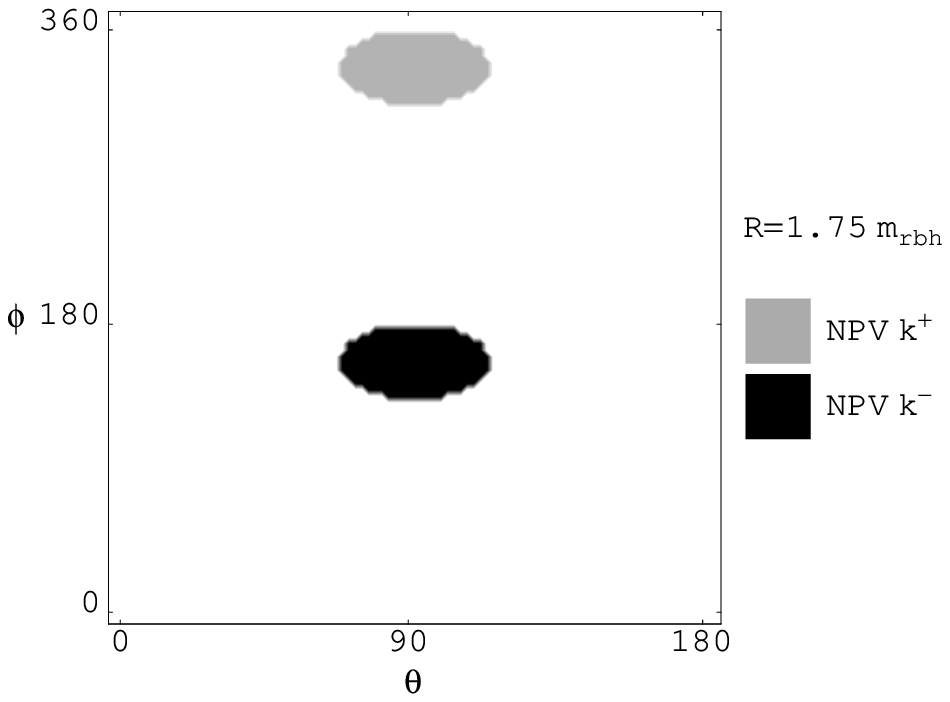,width=3.0in}\\
\epsfig{file=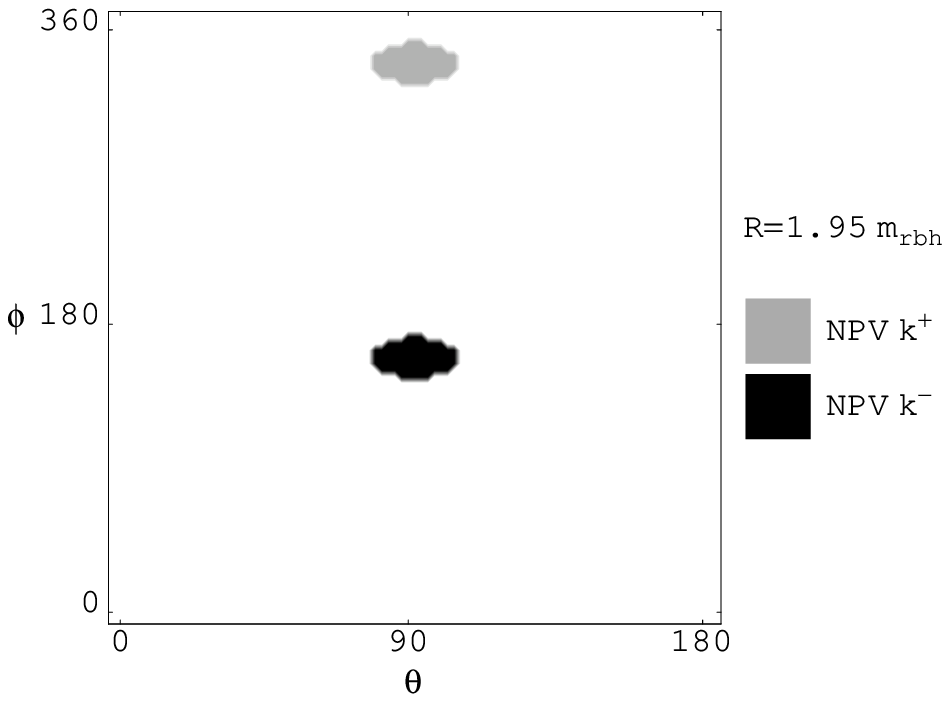,width=3.0in}
 \caption{\label{fig1}
The spherical polar coordinates $\theta \in [0^\circ, 180^\circ )$
and $\phi \in [0^\circ, 360^\circ )$ of the NPV wavevectors at the
points on the $x$ axis with $R = 1.55 \,\mrbh$, $R = 1.75 \,\mrbh$
and $R = 1.95 \,\mrbh$, for $\arbh = \sqrt{3/4}\,\mrbh $. Note
that the outer event horizon lies at $R_+ = 1.5\, \mrbh$ whereas
the stationary limit surface lies at $R_{S_+} = 2 \, \mrbh$ on the
$x$ axis.
 Orientations associated with the $k^+$ and $k^-$
wavevectors are shaded in gray and black, respectively.
 }
\end{figure}

\newpage

 \setcounter{figure}{1}
\begin{figure}[!ht]
\centering \psfull \epsfig{file=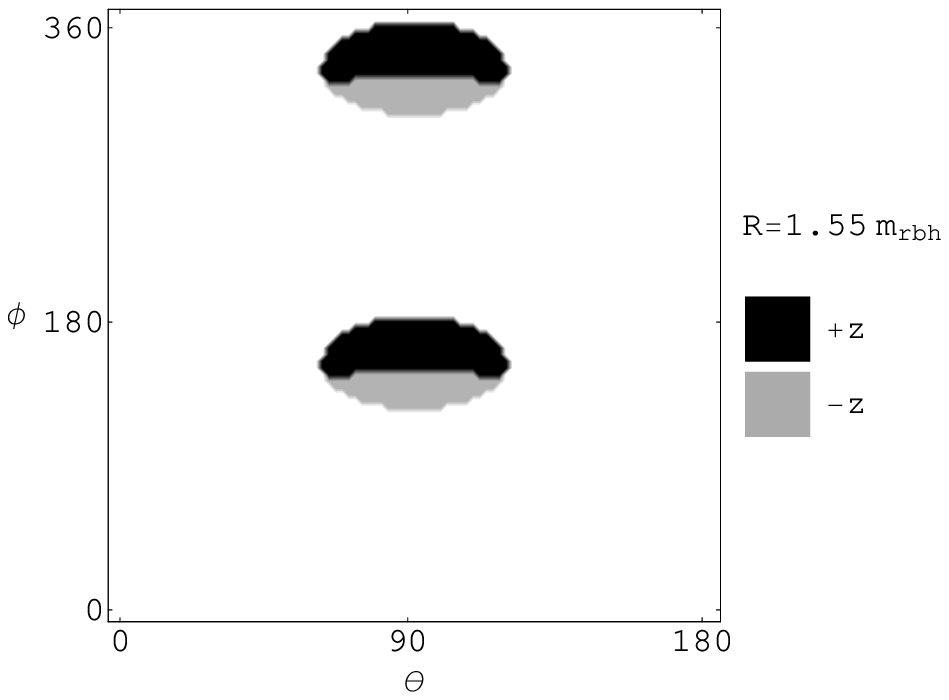,width=3.0in}\\
\epsfig{file=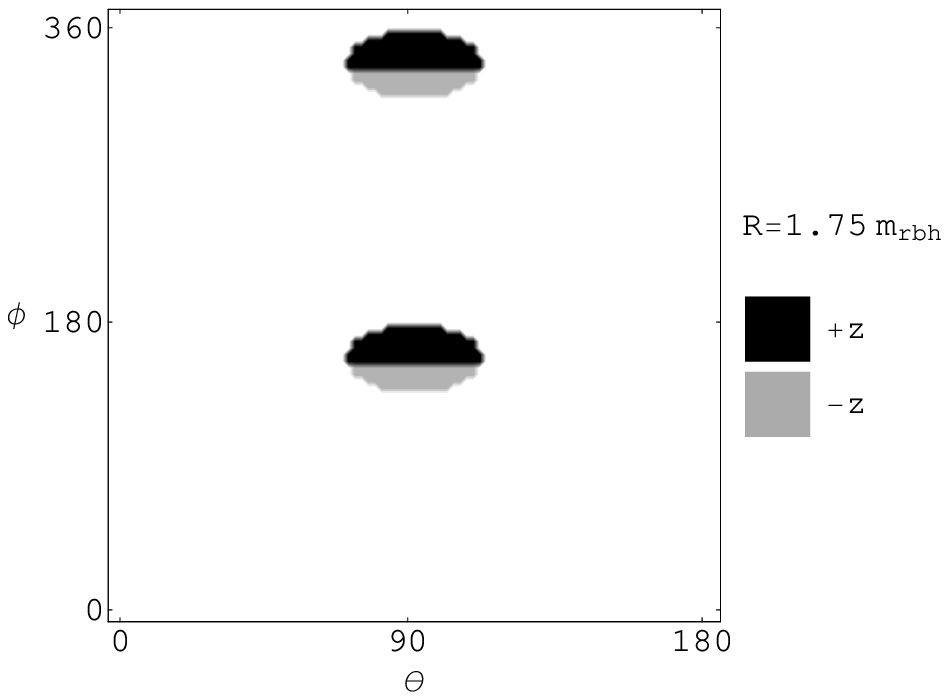,width=3.0in}\\
\epsfig{file=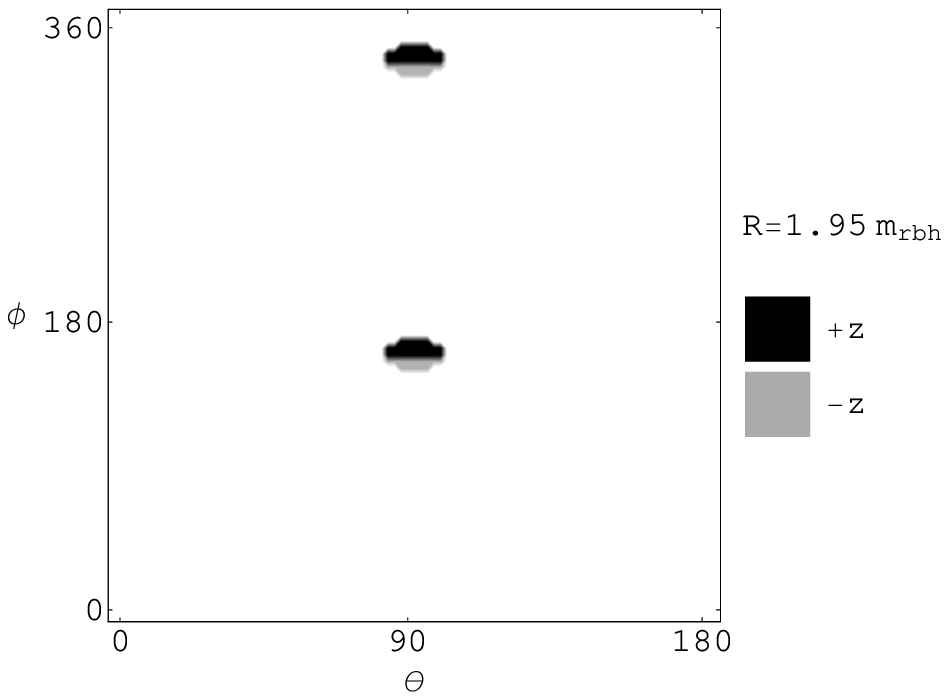,width=3.0in}
 \caption{\label{fig2}
 As figure~1 but with wavevector orientations associated with angular momentum
 in the positive $z$ direction  shaded in black and those associated
with angular momentum
 in the negative $z$ direction  shaded in
 gray.
 }
\end{figure}

\end{document}